\documentclass[final,3p,times,12pt]{elsarticle}
\usepackage{amssymb}
\usepackage{lineno}

\usepackage{graphicx}% Include figure files
\usepackage{amsmath}
\usepackage{subfig}
\usepackage{natbib}
 \def\vector#1{\mbox{\boldmath $#1$}}

\journal{Journal of Computational Physics}

\makeatletter
\def\@author#1{\g@addto@macro\elsauthors{\normalsize%
    \def\baselinestretch{1}%
    \upshape\authorsep#1\unskip\textsuperscript{%
      \ifx\@fnmark\@empty\else\unskip\sep\@fnmark\let\sep=,\fi
      \ifx\@corref\@empty\else\unskip\sep\@corref\let\sep=,\fi
      }%
    \def\authorsep{\unskip,\space}%
    \global\let\@fnmark\@empty
    \global\let\@corref\@empty  %% Added
    \global\let\sep\@empty}%
    \@eadauthor={#1}
}
\makeatother

\makeatletter
\def\ps@pprintTitle{%
   \let\@oddhead\@empty
   \let\@evenhead\@empty
   \let\@oddfoot\@empty
   \let\@evenfoot\@oddfoot
}

\begin{document}

\begin{frontmatter}


\title{Investigation of the Energy Shielding of Kidney Stones by Cavitation Bubble Clouds during Burst Wave Lithotripsy}
\author[Caltech]{Kazuki Maeda\corref{cor1}}
%\author{Kazuki Maeda\corref{cor1}\fnref{label2}}
\ead{maeda@caltech.edu}
\cortext[cor1]{Corresponding author}

\author[UW1,UW2]{Adam D. Maxwell}
\author[UW1]{Wayne Kreider}
\author[Caltech]{Tim Colonius}
\author[UW1,UW2]{Michael R. Bailey}

%\title{Symmetry reduction of the volume-averaged Navier-Stokes equation for
%simulations of dispersed multiphase flows}


%\author{Kazuki Maeda$^*$ and Tim Colonius}

\address[Caltech]{Division of Engineering and Applied Science, California Institute of Technology\\
1200 East California Boulevard, Pasadena, CA 91125, USA}
\address[UW1]{Center for Industrial and Medical Ultrasound, Applied Physics Laboratory, University of Washington 1013 NE 40th St, Seattle, WA 98105, USA}
\address[UW2]{Department of Urology, University of Washington School of Medicine, 1959 NE Pacific St, Seattle, WA 98195, USA}

\begin{abstract}
We conduct experiments and numerical simulations of the dynamics of bubble clouds nucleated on the surface of an epoxy cylindrical stone model during burst wave lithotripsy (BWL).  In the experiment, the bubble clouds are visualized and bubble-scattered acoustics are measured. In the numerical simulation, we combine methods for modeling compressible multicomponent flows to capture complex interactions among cavitation bubbles, the stone, and the burst wave. Quantitative agreement is confirmed between results of the experiment and the simulation. We observe and quantify a significant shielding of incident wave energy by the bubble clouds.  The magnitude of shielding reaches up to 80\% of the total acoustic energy of the incoming burst wave, suggesting a potential loss of efficacy of stone comminution. We further discovered a strong linear correlation between the magnitude of the energy shielding and the amplitude of the bubble-scattered acoustics, independent of the initial size and the void fraction of the bubble cloud within a range addressed in the simulation.  This correlation could provide for real-time monitoring of cavitation activity in BWL.
\end{abstract}


\end{frontmatter}

%%\linenumbers

\section{Introduction}
\label{S:1}
Cavitation bubble clouds nucleated in the human body during the passage of tensile components of ultrasound pulses are of critical importance for the safety and efficacy of the treatment of lithotripsy\cite{Pishchalnikov03,Ikeda06}. Shock wave lithotripsy (SWL) uses a shock wave form of ultrasound pulses with a peak amplitude of $O(10-100)$ MPa. The large amplitude of the pulses followed by a long tensile tail tends to results in formation of a large bubble cluster that can violently collapse to cause tissue injury, which has been seen as a major disadvantage of SWL\cite{McAteer05,Bailey06}. Burst wave lithotripsy (BWL) is a newly proposed alternative to SWL that uses pulses of a continuous form of pressure wave (burst wave) with an amplitude of $O(1-10)$ MPa and a frequency of $O(100)$ kHz for stone comminution\cite{Maxwell05}. Due to the low peak amplitude of the wave, resulting cloud cavitation in BWL is expected to involve smaller bubbles and less violent bubble dynamics. Recent in vitro experiments show that cavitation bubble clouds with a size of $O(1)$ mm are formed during a passage of the burst wave\cite{Maeda15}. Such bubble clouds can scatter and absorb the incoming burst wave to cause energy shielding of a nearby kidney stone. The magnitude of this shielding has not been quantitatively analyzed, despite its importance for the efficacy of stone comminution.
In the present study, we quantify the energy shielding by bubble clouds nucleated on the surface of an epoxy stone model during the passage of a burst wave with an amplitude of 7 MPa and a frequency of 340 kHz generated by a focused ultrasound transducer in water through experimental measurements and numerical simulation. In the experiment, we visualize the evolution of bubble clouds using a high-speed camera and measure the back-scattered acoustics from the bubbles with transducer array elements. In the simulation, we combine novel numerical methods for modeling compressible multiphase flows to capture complex interactions among bubbles, the stone, and the burst wave. Simulated evolution of the bubble cloud and bubble scattered acoustics quantitatively agree with the results of the experiment. We vary the initial void fraction and size of bubble cloud to assess the magnitude of the shielding as well as its correlation with the bubble-scattered acoustics. Results of the study indicate that the magnitude of the shielding reaches up to 80\% of the total acoustic energy of the incident burst wave. We further discovered a strong linear correlation between the magnitude of the shielding and the amplitude of the back-scattered acoustics, independent of the initial condition of the cloud. The results of the study can be used to further identify appropriate parameters and improve the efficacy of stone comminution in BWL.

\section{Experimental and numerical setups}
\label{section:setup}
\begin{figure}
  \center
  \includegraphics[width=140mm,trim=0 0 0 0, clip]{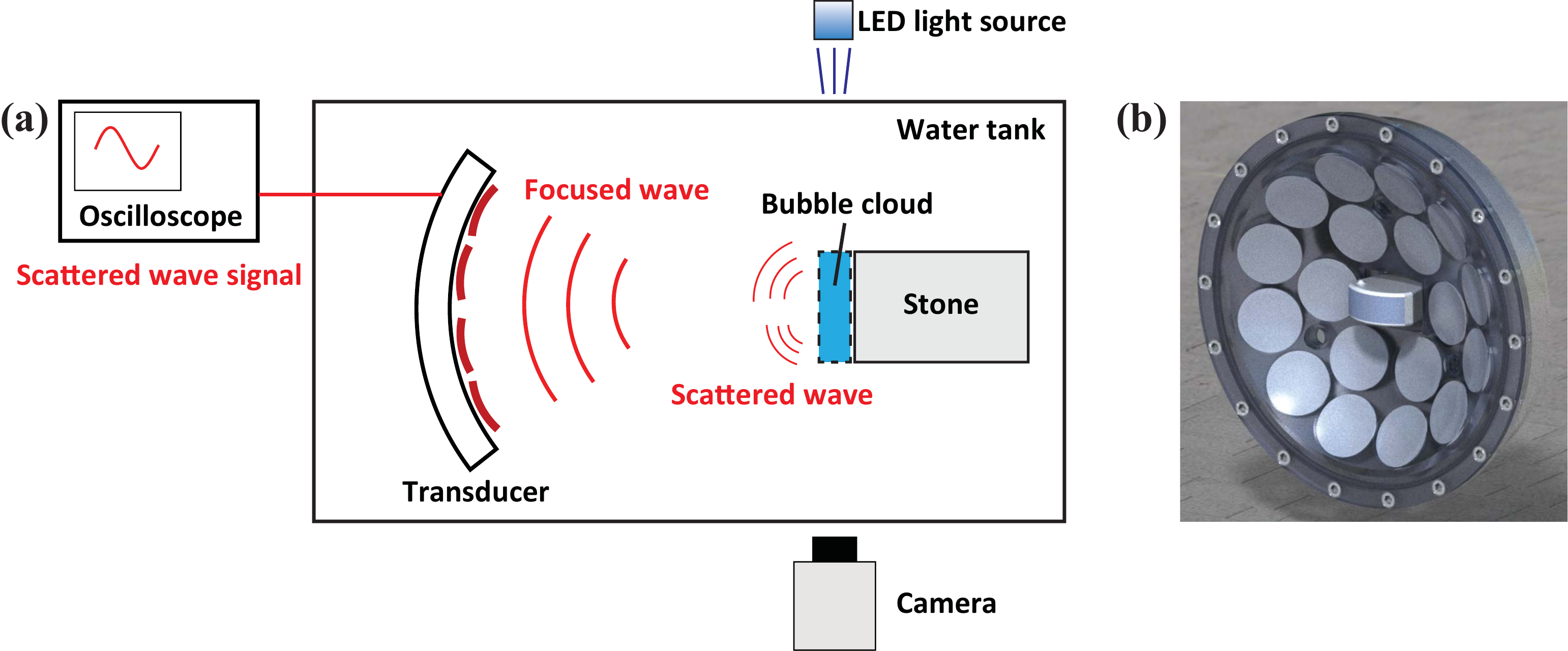}
%  \end{subfigmatrix}
  \caption{(a) Schematic of the experimental setup. (b) Multi-element array medical transducer.}
   \label{fig:1} 
\end{figure}
Fig. \ref{fig:1}a shows a schematic of the experimental setup. We generate pulses of 10 cycles of a sinusoidal pressure wave with a frequency of 340 kHz and a peak focal pressure of 7 MPa from a multi-element array focused transducer with an aperture of 180 mm and focal length of 150 mm (fig. 1b) toward a cylindrical shape of epoxy stone model with its axis aligned on the acoustic axis of the transducer. The pulse repetition frequency is 100 Hz. The focal point of the transducer is located at the center of the stone. The height and radius of the stone are 10 and 6.25 mm, respectively. A high-speed camera captures a thin layer of bubbles forming a cloud nucleated on the proximal base of the stone. We concurrently sample the back-scattered acoustics from the bubble cloud and the stone by using the transducer array elements.
In simulations, we formulate the dynamics of the multi-component mixture using the compressible, multi-component, Navier-Stoke equation. We model the stone as an elastic solid with zero shear modulus with a density of 1200 kgm$^{-3}$ and a longitudinal sound speed of 2500 ms$^{-1}$. The coupled dynamics of the stone and the surrounding water are modeled using an interface capturing method\cite{Coralic14}. We model the transducer as a portion of a spherical surface with an aperture of 60 mm and a radius of 50 mm with its center located at the origin. We generate burst waves at the modeled surface by using a source term approach\cite{Maeda17}. The governing equations are discretized on an axisymmetric grid and integrated using a fifth-order, finite volume WENO scheme. The grid size is uniform with a radial and axial width of 100 $\mu$m in the stone-wave interaction region. For modeling bubbles, we utilize an Eulerian-Lagrangian method \cite{Maeda17b}. We initially distribute spherical bubble nuclei with a radius of 10 $\mu$m in the cylindrical region that faces the proximal base of the stone, then track the radial evolution of the bubbles at the sub-grid scale and resolve the bubble-scattered acoustics on the grid. For the parametric study, we run simulations varying the thickness of the bubbly layer, $h$, and the initial void fraction, $\beta_0$, of the cylindrical cloud, within ranges of $h\in[0.25, 1.5]$ mm and $\beta_0\in[1.0, 8.0]\times10^{-5}$, respectively. We capture the evolution of the bubbles and sample the back-scattered acoustics at points on the acoustic source surface corresponding to the locations of the center of the transducer array elements.

\section{Results and discussion}
\begin{figure}
  \center
% \begin{subfigmatrix}{1}
  \subfloat[]{\includegraphics[width=80mm,trim=0 -30 0 0, clip]{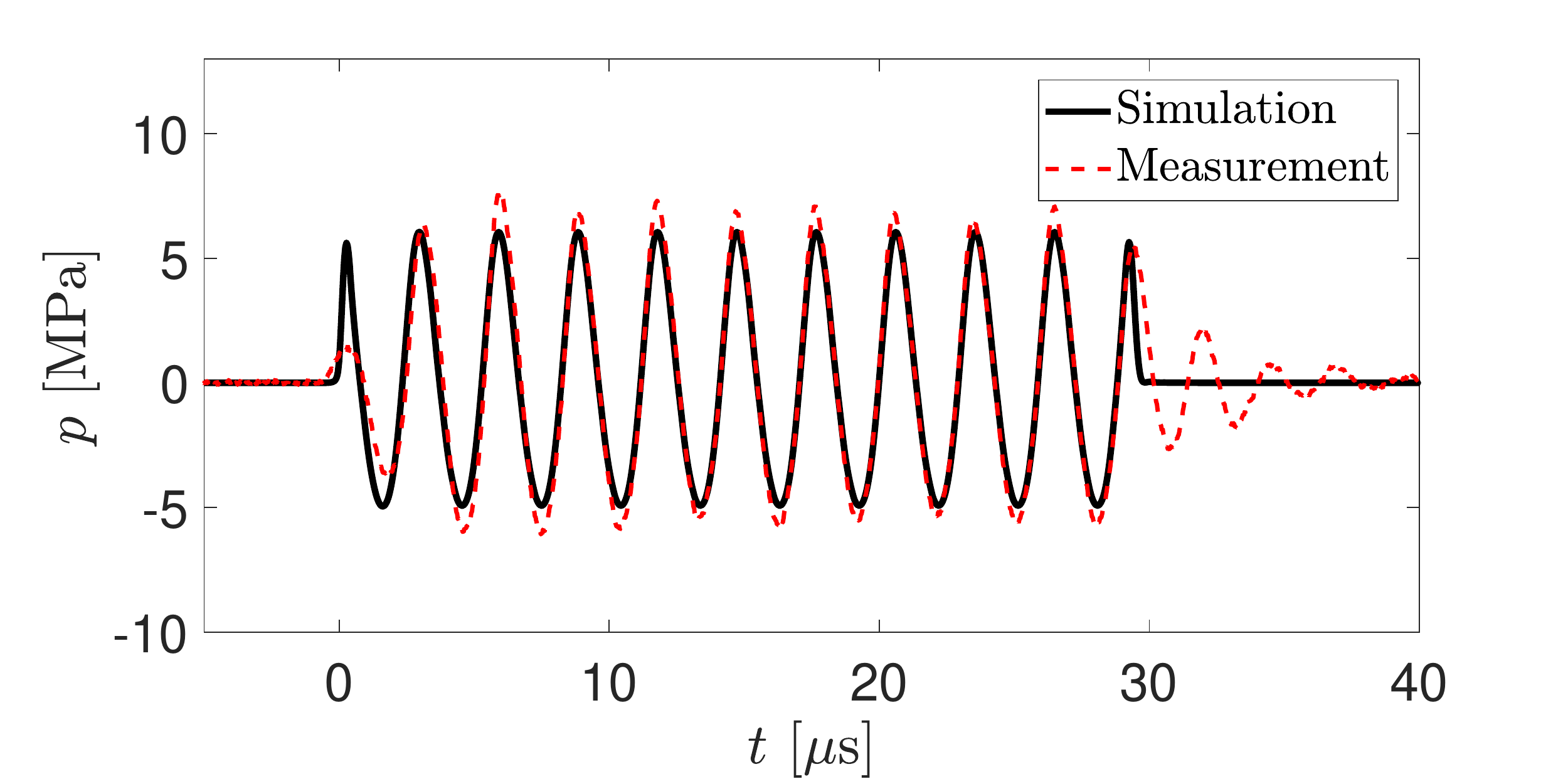}}
  \subfloat[]{\includegraphics[width=60mm,trim=0 0 0 0, clip]{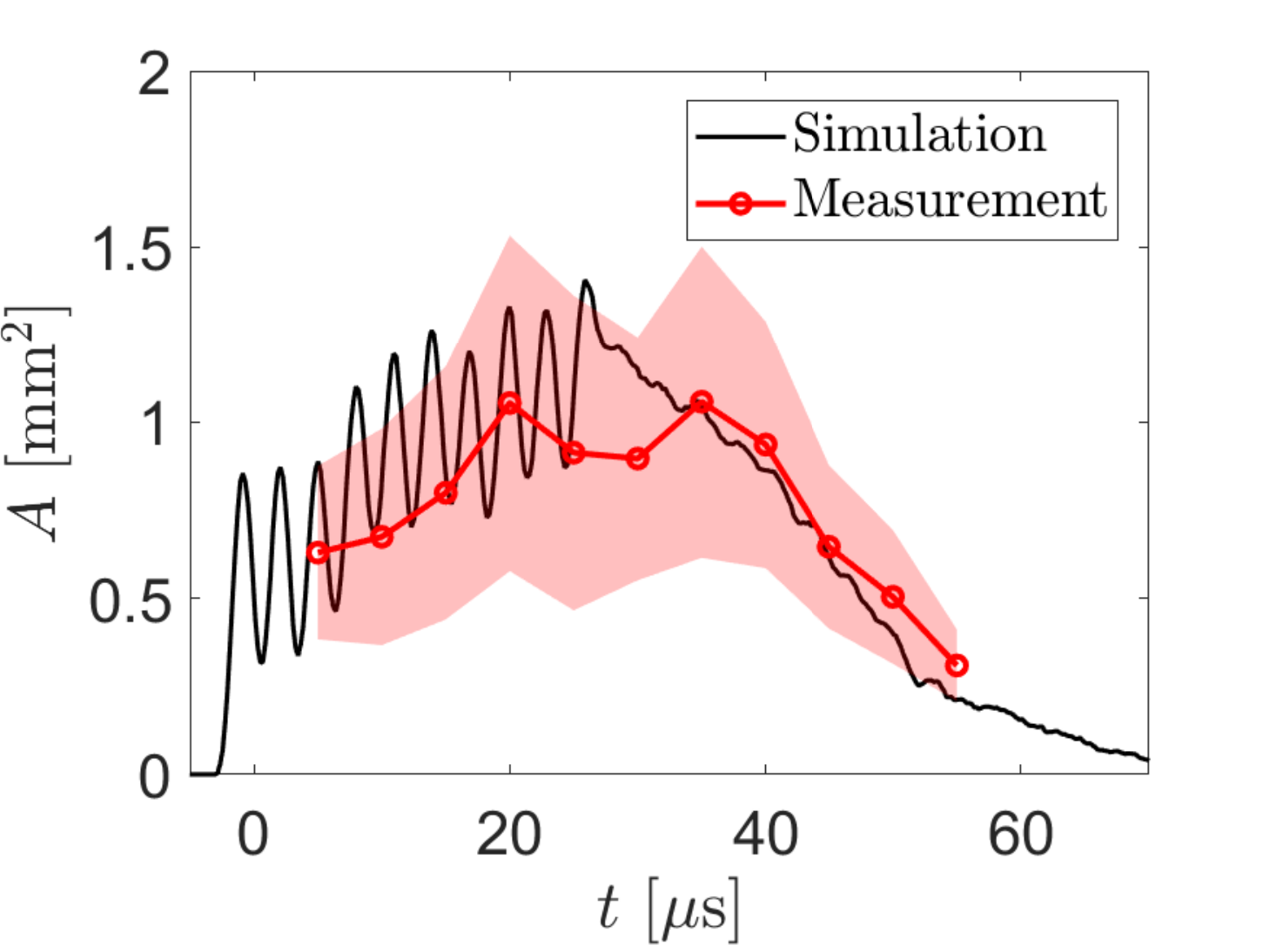}}\\
  \subfloat[]{\includegraphics[width=100mm,trim=0 0 0 0, clip]{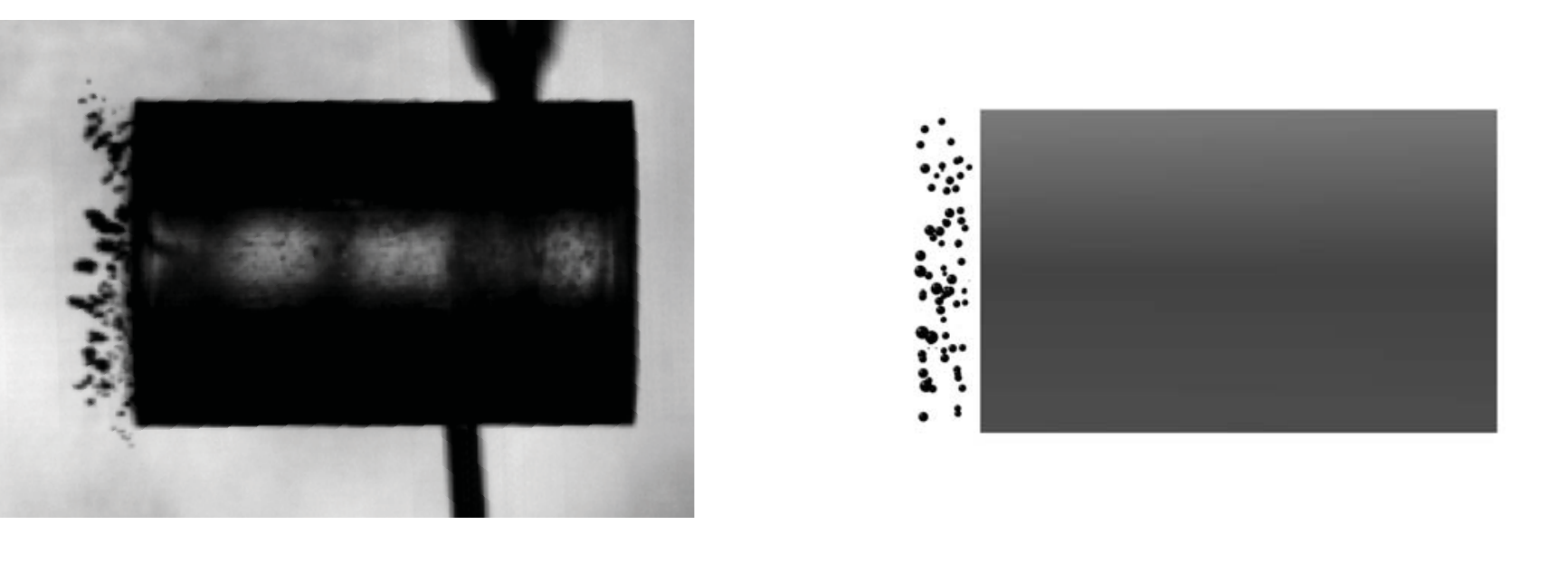}}
%  \end{subfigmatrix}
  \caption{Evolution of the focal pressure. (b) Evolution of the projected area of bubble cloud. Result of the simulation with $h = 1$ mm and $\beta_0 = 1.0\times10^{-5}$ is shown. (c) Images of representative bubble clouds obtained at $t = 20$ $\mu$s from the experiment and the simulation.
   }
   \label{fig:2} 
\end{figure}
In fig. \ref{fig:2}a we compare the evolution of the focal pressure during the passage of the burst wave obtained from the experiment and a simulation without the stone. The head of the wave arrives at the focal point at $t = 0$. The simulated waveform and the peak pressure agree well with those of the measurement, except for the small oscillations due to ring down after $t = 30$ $\mu$s. In fig. \ref{fig:2}b we compare the evolution of the area occupied by the bubbles projected on a 2D plane, $A$ $\mathrm{mm}^2$, obtained in the experimental visualization and the simulation with $h = 1.0$ mm and $\beta_0 = 1.0\times10^{-5}$. Each data point of the measurement is obtained by averaging 440 independent realizations of the bubble clouds. The shaded area corresponds to the standard deviation resulting from the randomness in the clouds. The result of the simulation shows that the cloud experiences a transient growth with rapid oscillations during the passage of the wave, up to about $t =30$ $\mu$s, followed by a smooth decay after the passage of the wave. The result of the measurement cannot capture the oscillations, even though they may be present in the experiments, as the present sampling frequency was lower than that of the oscillation, while its growing trend until around at $t = 30$ $\mu$s and decay after the passage of the wave agree well with the result of the simulation. Fig. 2c shows representative images of bubble cloud obtained from the experimental visualization and the simulation at $t = 20$ $\mu$s. A layer of dispersed bubbles is present on the proximal base of the stone in both images.
\begin{figure}
  \center
  \includegraphics[width=150mm,trim=0 0 0 0, clip]{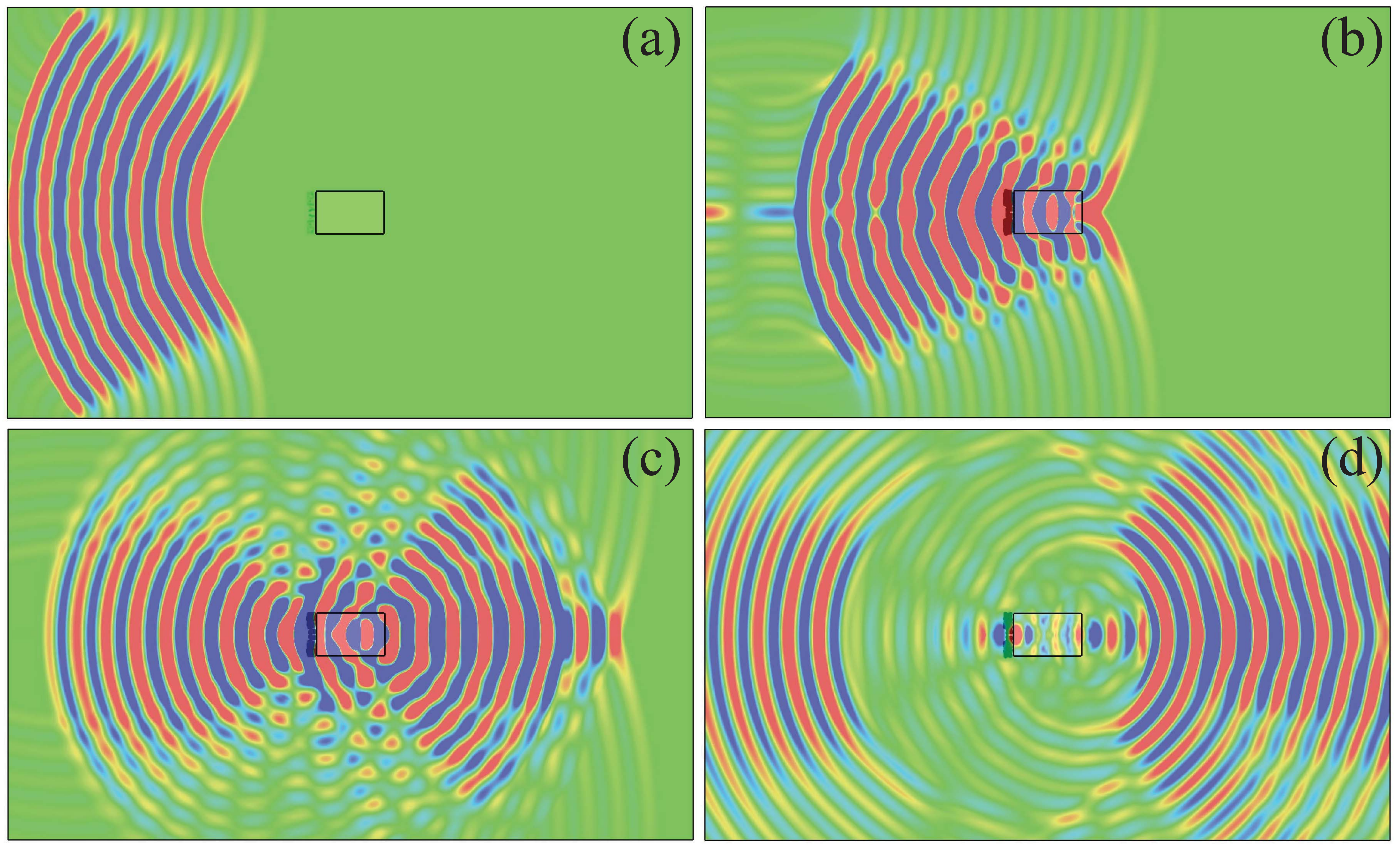}
%  \end{subfigmatrix}
  \caption{Snapshots of the pressure contour on the axial-plane during the simulation with $h = 1.0$ mm and $\beta_0 = 1.0\times10^{-5}$ at (a) $t = 10$, (b) 20, (c) 30, and (d) 40 $\mu$s. Stone and the void fraction of bubbles are indicated with black line and shading, respectively.}
   \label{fig:3} 
\end{figure}
Fig. \ref{fig:3} shows pressure contours at different instants in time during the simulation. A burst wave is generated and focused toward the stone (fig. \ref{fig:3}a). A bubble cloud is excited during the passage of the wave (fig. \ref{fig:3}b). A portion of the wave is scattered back from the bubbles and the stone, and the rest of the wave transmits through the stone and is scattered forward (fig. \ref{fig:3}b). After the passage of the incident wave, weak reverberation of the pressure wave is observed in the stone (fig. \ref{fig:3}d). Strong collapse of bubble cloud is not observed. This may be due to the small void fraction and size of the cloud.
In order to quantify the amplitude of the back-scattered acoustics, we compute an imaging functional using the acoustic signals sampled in both the experiment and the simulation. The imaging functional at coordinate $\vector{z}$ is defined as:
\begin{equation}
I^B(\vector{z})=\sum^{N_{\mathrm{sensor}}}_{j,l=1}C(\Delta t(\vector{z},\vector{x}_j,\vector{x}_l),\vector{x}_j,\vector{x}_l)
\end{equation}
where $C$ is the cross correlation of the signals obtained at coordinates $x_j$ and $x_l$. $\Delta$ is the difference between the acoustic travel time from $\vector{z}$ to $\vector{x}_l$ and that from $\vector{z}$ to $\vector{x}_j$. Details of the algorithms to obtain the functional can be found in \cite{Garnier09}.
\begin{figure}
  \center
% \begin{subfigmatrix}{1}
  \subfloat[]{\includegraphics[width=52mm,trim=0 -4 0 0, clip]{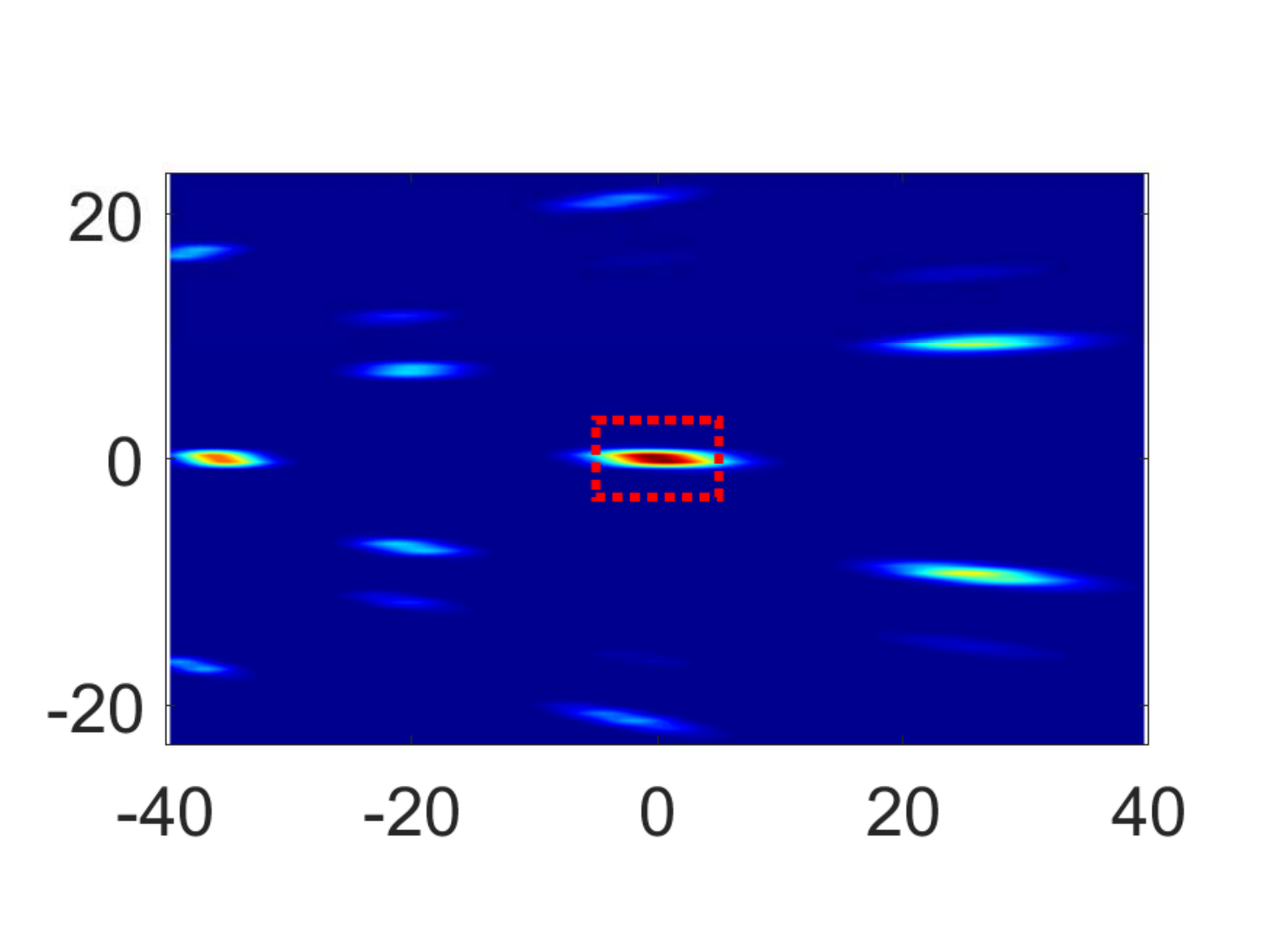}}
  \subfloat[]{\includegraphics[width=80mm,trim=0 50 0 0, clip]{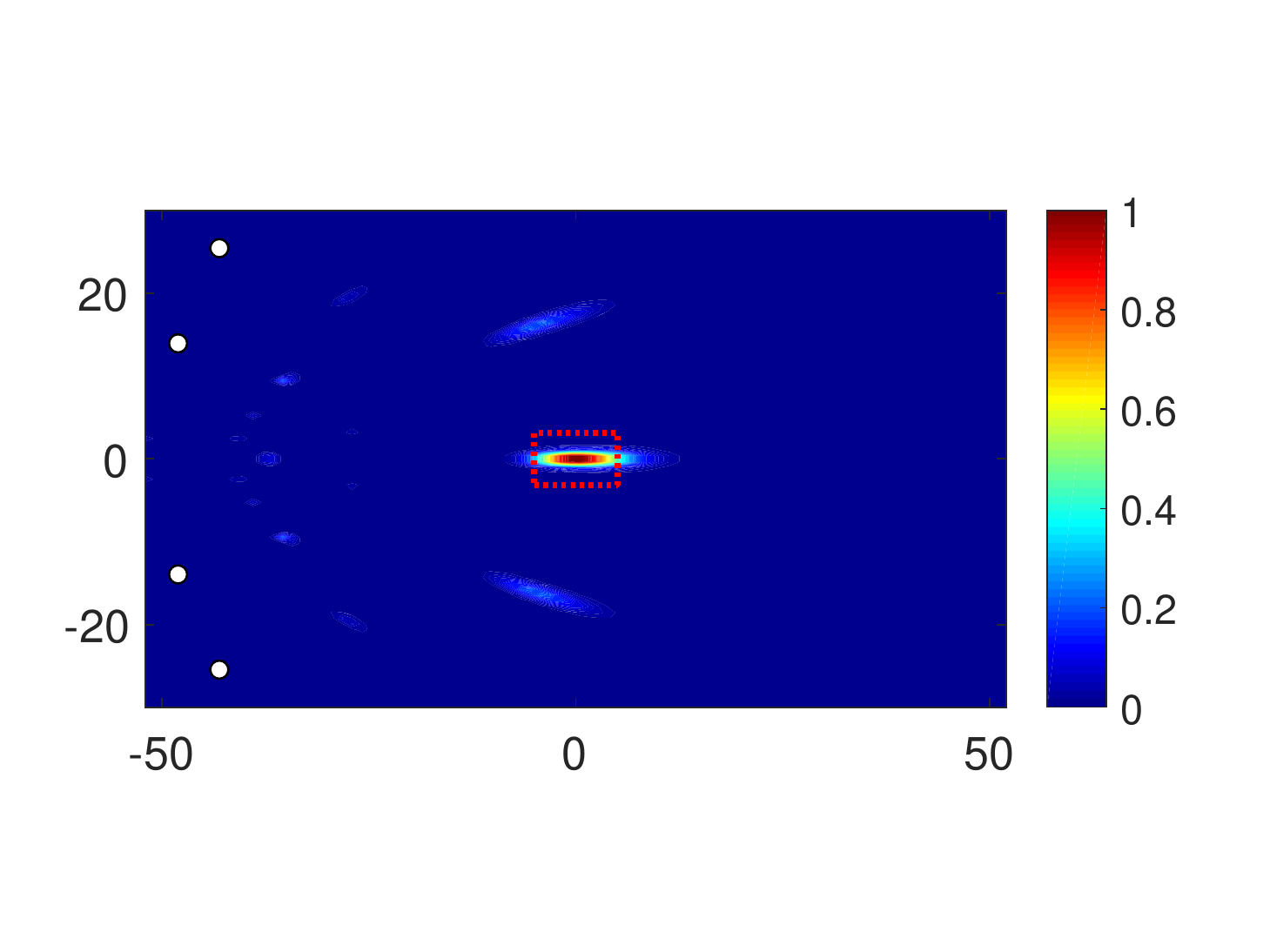}}\\
  \subfloat[]{\includegraphics[width=70mm,trim=0 0 0 0, clip]{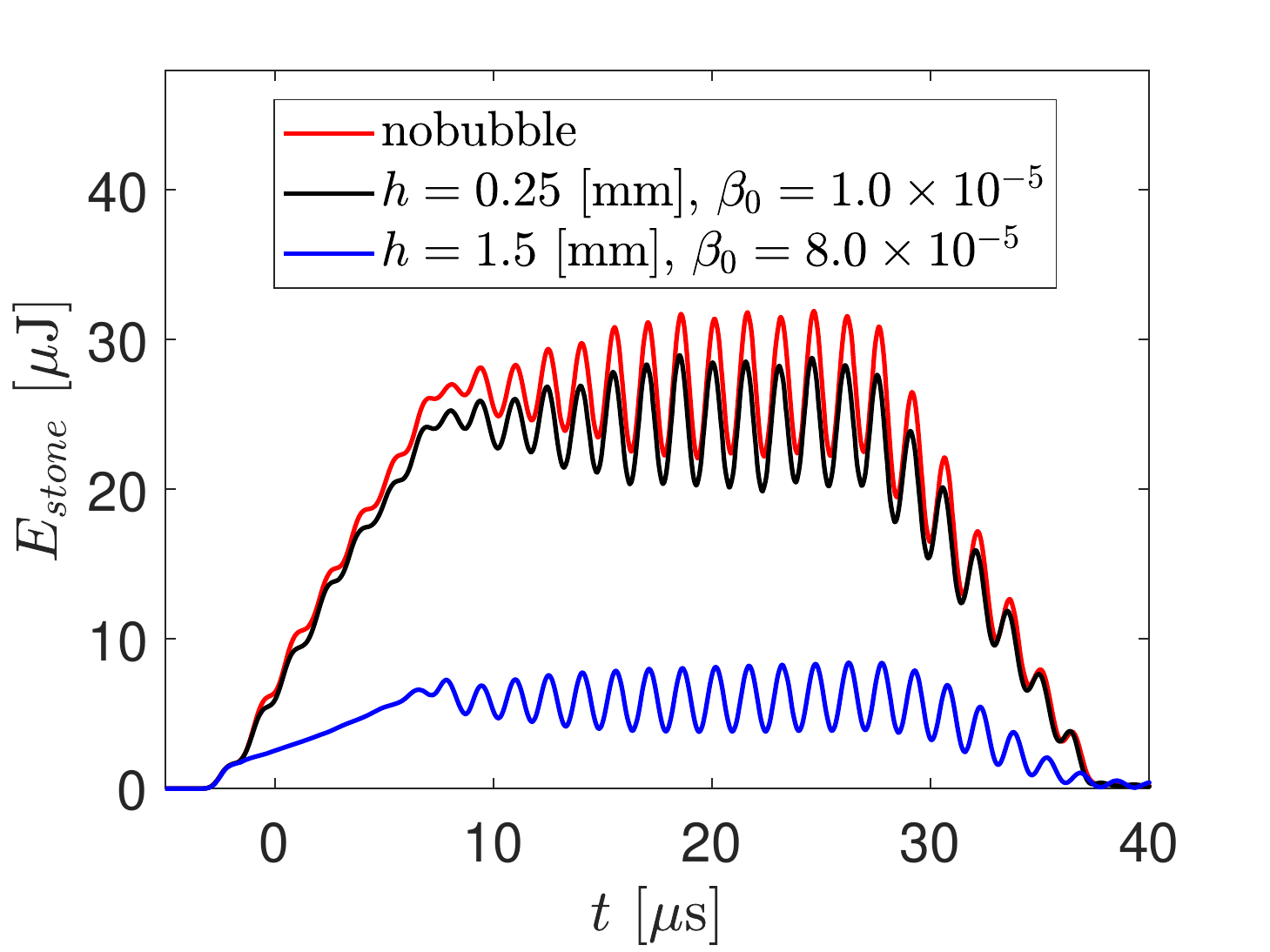}}
  \subfloat[]{\includegraphics[width=70mm,trim=0 0 0 0, clip]{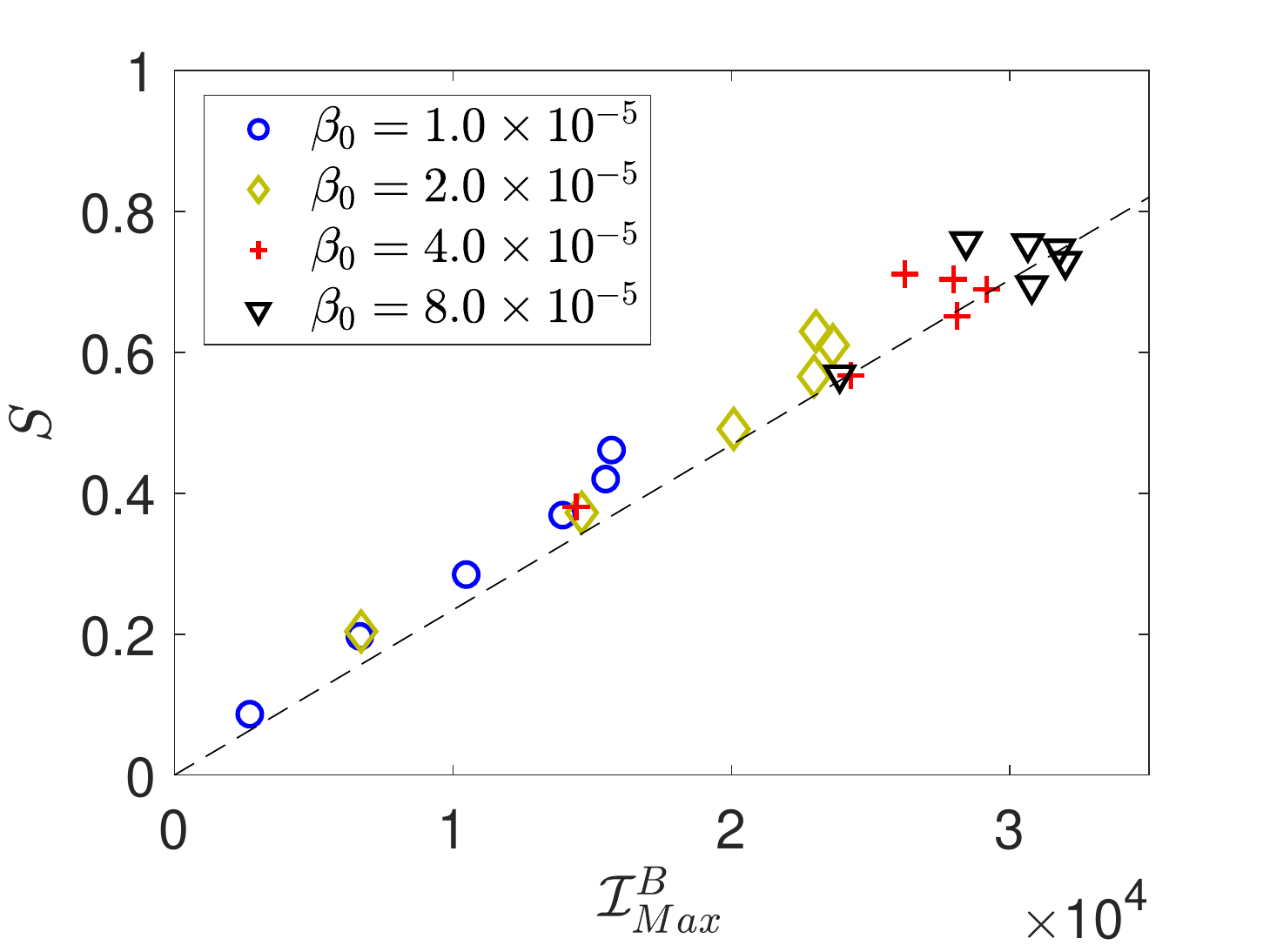}}
%  \end{subfigmatrix}
  \caption{(a) Normalized contour of the imaging functional obtained from the experiment. Stone is indicated by a dotted line. The length unit is mm. (b) That obtained from the simulation. Sensor locations are indicated by white dots. (c) Evolution of the acoustic energy in the stone. (d) Correlation between the shielding factor and the maximum value of the imaging functional.}
   \label{fig:4} 
\end{figure}
Fig. \ref{fig:4}a and \ref{fig:4}b show normalized contours of the imaging functional obtained in the experiment and the simulation. Coordinates with a large value of the imaging functional correspond to estimated locations of the acoustic source.
In both plots, a region of a large value of the imaging functional is localized within the area of the stone.
This result suggests that the acoustics scattered from the stone (and the bubbles) are captured in both the experiment and the simulation, and their magnitude can be quantified by the imaging functional. Fig. \ref{fig:4}c shows the evolution of the acoustic energy in the stone during three distinct simulation cases: without bubbles; with a thin and dilute cloud ($h = 0.25$ mm and $\beta_0 = 1.0\times10^{-5}$); with a thick and dense cloud ($h = 1.5$ mm and $\beta_0 = 8.0\times10^{-5}$). In all the cases, the energy steadily grows after arrival of the wave until around at $t = 10$ $\mu$s, then oscillates around a constant value during the passage of the wave until around at $t = 30$ $\mu$s. After the passage of the wave, the energy steadily decays and reaches
zero around at $t = 40$ $\mu$s. The energy is highest for all $t$ when there are no bubbles. The energy level is slightly decreased with the thin and dilute cloud, while it is drastically reduced with the thick and dense cloud. This result indicates that the energy shielding by bubbles is enhanced by increasing the thickness and/or the void fraction of the cloud. To quantify the correlation between the shielding and the scattered acoustics, in fig \ref{fig:4}d we plot the shielding factor $S$ as a function of the maximum value of the imaging functional, $I^B_{\mathrm{Max}}$. The shielding factor is defined as
\begin{equation}
S=1-\frac{P}{P_{\mathrm{ref}}},
\end{equation}
where $P$ is the total work done by the acoustic energy to the stone during each simulation: $P=\int E_{\mathrm{stone}}dt$. $P_{\mathrm{ref}}$ is the reference value of $P$ obtained in the case without bubbles. Note that $S=0$ and $S=1$ indicates no shielding and perfect shielding (no energy transmission into the stone), respectively. The maximum value of the shielding factor in the plot is approximately 0.8, indicating that up to 80\% of the total energy of the incident burst wave can be absorbed/scattered by bubbles that otherwise transmits into the stone. Interestingly, the plot also indicates a strong linear correlation between the shielding factor and the maximum imaging functional within the global range of the shielding factor, independent of the initial condition of bubbles. This correlation suggests that the scattered acoustics can be directly used to estimate the magnitude of the shielding regardless of the bubble dynamics, at least within the range of parameters of bubble cloud addressed in the present study.

\section{Conclusion}
We quantified the energy shielding of kidney stones by a bubble cloud nucleated on the stone surface during the passage of a burst wave through a combined experimental and numerical approach. Simulated evolution of the bubble cloud and the bubble-scattered acoustics using a compressible multicomponent flow solver showed quantitative agreement with the results of high-speed photography and acoustic measurements. Results of the simulation revealed that the magnitude of the energy shielding by a thin layer of bubble cloud can reach up to 80\% of the total energy of the burst wave, indicating a large potential loss of efficacy in the treatment of BWL due to cloud cavitation. Furthermore, we discovered a strong linear correlation between the magnitude of the shielding and the amplitude of the back-scattered acoustics. This correlation could be used, for example, to monitor the magnitude of the shielding in a human body by using ultrasound imaging in real time during the treatment of BWL, and adjust parameters to minimize shielding during a procedure. Future work can include further detailed analysis of the bubble-scattered acoustics as well as assessment of the effect of the amplitude and frequency of the burst wave on the energy shielding.

\section*{Acknowledgments}
K.M would like to acknowledge the Funai Foundation for Information Technology, for the Overseas Scholarship. This
work was supported by the National Institutes of Health under grants P01-DK043881 and K01-DK104854. The
simulations presented here utilized the Extreme Science and Engineering Discovery Environment, which is supported
by the National Science Foundation grant number CTS120005.

%\section*{Citations}
\bibliographystyle{unsrt}
\bibliography{sample}






% \end{thebibliography}

\end{document}